# Dynamic Properties of Water inside Graphene Oxide Membranes

One-Sun Lee*

Qatar Environment and Energy Research Institute, Hamad Bin Khalifa University,

P.O. Box 5825, Doha, Qatar

*Corresponding Author

Electronic Mail: onesun2@gmail.com

Phone: +974 3366 6845

**Abstract**

Separation of salt ions from seawater using graphene oxide (GO) membrane is an emerging desalination method. However, the structure and dynamics of water molecules and ions inside GO membrane are not known, and this has proven to be a hindrance in understanding how this system functions. Here, we investigate the dynamic behavior of water, sodium, and chloride ions inside GO membrane using a computational approach. First, we developed four different models of GO membrane with different distances between GO sheets (7, 9, 11, and 13 Å), and performed molecular dynamics (MD) simulations of water inside each membrane for 20 ns. With the analysis of the mean squared displacement of water and the Einstein relation equation for diffusion, we measured the diffusion coefficient of water in each membrane. We found that the diffusion coefficient of water inside the GO membrane decreased as the distance between sheets ($d$) decreased. The measured diffusion coefficients (in unit of $\times 10^{-5}$ cm$^2$/sec) of water are 0.03 ($d$ = 7 Å), 0.91 ($d$ = 9 Å), 1.18 ($d$ = 11 Å), and 1.49 ($d$ = 13 Å), whereas it is 2.49 (simulation) or 2.3 (experiment) for bulk water. Second, for analyzing the desalination performance of each membrane, we also calculated the free energy profile of penetration of water, sodium, and chloride ions via the interlayer of the membrane when $d$ = 7, 9, 11, and 13 Å. We employed the steered MD simulations for guiding the target molecule and measured the free energy profile with the Jarzynski equation using eight independent trajectories of each target molecule. The free energy profile of penetration revealed that the optimum interlayer distance between GO sheets for desalination is 9 ~ 11 Å. With 9 and 11 Å interlayer distance between GO sheets, the free energy barrier of membrane penetration of water is negligible, where it is 2 ~ 4 kcal/mol for sodium or chloride ions. We found that the membrane with 13 Å interlayer distance does not show an energy barrier for sodium and chloride ions as well as water. We also found that 7 Å is the minimum distance between GO sheets that could accommodate water molecules inside GO membrane, and the membrane with a 7 Å interlayer distance has a high rejection energy barrier for sodium and chloride ions. However, water molecules confined between 7 Å interlayer distance GO sheets show low mobility because of the interaction with GO sheets such as hydrogen bonding. We believe that our simulation results would be a significant contribution to designing a new GO-based membrane for desalination.

# I. Introduction

The separation of salt ions from seawater using graphene-based membrane is an emerging desalination method.[1-3] According to previous simulation and theoretical works, graphene-based membranes may achieve water permeability up to 1000 times greater than that of commercial seawater membranes.[4-5] Therefore, there have been many experimental approaches for developing better graphene-based membranes for desalination with the support of computational simulations. However, most of the simulation of desalination adapts only one or few sheets of graphene for modeling the membrane. With this minimal model, the information of dynamic properties of water and ions is restricted and this has proven to be a hindrance in understanding how this system functions since the properties of confined water in nano-scale pores or channels of membranes are expected to be different from those of the bulk liquid.

Molecular dynamics (MD) simulations offer information that is hardly accessible by experiments, and they fill the gap between theories and experiments.[6-17] Here we developed a model membrane system based on the experimentally suggested[1-3] stacked graphene model (Figure 1 (A)). To develop a large system of membranes, we adapted the periodic boundary conditions to our model.[18] In periodic boundary conditions, when a molecule passes through one side of the main unit cell, it comes back on the opposite side, and each individual particle in the main simulation cell interacts with the closet image (Figure 1 (B), (C), (D), and (E)). We performed MD simulations of water inside this stacked graphene-based membrane. By varying the density of water inside the membrane, we could investigate the dynamic property changes of water. We found that the translation diffusion coefficient of water increases as the density of water decreases. We also found that hydrogen bond lifetime increases as the density of water decreases. Also, we measured the free energy profile of water and ions penetration between two graphene sheets using steered molecular dynamics simulations and the analysis with the Jarzynski equation.[19-21] The free energy profile of penetration revealed that the optimum interlayer distance between graphene sheets for desalination is ~10 Å. We believe that our simulation results would be a significant contribution to designing a new graphene-based membrane for desalination.

## II. Computational Details

**GO Membrane**

The GO membrane is composed of four layers of GO, and each layer is composed of four rectangular GO with the dimension of 20 Å × 20 Å (Figure 1 (B)). The four GO membrane systems were prepared with different interlayer distances $d$ (7, 9, 11, and 13 Å). The hydroxyl and epoxy groups are randomly functionalized on the surface of bare graphene to generate a GO segment. Based on previous experiments, we set the ratio of 5:1 for carbon versus oxygen and 3:2 for the hydroxyl versus epoxy groups.[22] The SPC/E water model is adapted for all simulations. The number of water molecules in each membrane was determined by comparing the MD simulations with NPT and NVT ensembles. We performed several MD simulations with NPT ensembles by varying the number of water molecules. Then we monitored the volume changes of each system and chose the one that has volume fluctuation of less than 1% comparing with the MD simulation with the NVT ensemble. The summary of the systems is shown in Table 1.

**Force Field Parameters**

The energy function used in the simulations has the form

$$U(b,\theta,\chi,\varphi,r_{ij}) = \sum_{bonds} K_b(b-b_0)^2 + \sum_{angle} K_\theta(\theta-\theta_0)^2 + \sum_{dihedral} K_\chi(1+\cos(n\chi-\delta)) + \sum_{impropers} K_{imp}(\varphi-\varphi_0)^2$$

$$+ \sum_{nonbond} \varepsilon_{ij}\left[\left(\frac{R_{ij}^{min}}{r_{ij}}\right)^{2} - \left(\frac{R_{ij}^{min}}{r_{ij}}\right)^{6}\right] + \frac{q_i q_j}{\varepsilon_0 r_{ij}} \quad (1)$$

where $K_b$, $K_\theta$, $K_\chi$, and $K_{imp}$ are bond, angle, dihedral angle, and improper dihedral angle force constants, respectively; $b$, $\theta$, $\chi$, and $\varphi$ are bond length, angle, dihedral angle, and improper torsion angle, respectively, with the subscript 0 representing the equilibrium values for the individual terms, and $n$ determining the periodicity of the dihedral potential in the interval $[0,2\pi]$. Lennard-Jones (LJ) 6-12 and Coulombic terms contribute to the nonbonded interaction where $\varepsilon_{ij}$ is the LJ well depth, $R_{ij}^{min}$ is the distance at the LJ minimum, $q_i$ is the partial atomic charge, $\varepsilon_0$ is the effective dielectric constant, and $r_{ij}$ is the distance between atoms *i* and *j*. The $\varepsilon_{ij}$ values were obtained via the geometric mean ($\varepsilon_{ij} = \sqrt{\varepsilon_{ii}\varepsilon_{jj}}$) and $R_{ij}^{min}$ via the arithmetic mean ($R_{ij}^{min} = \dfrac{R_i^{min} + R_j^{min}}{2}$).

**Molecular Dynamics Simulations**

The system is equilibrated for 1000 steps using the conjugate gradient method and followed by 10 ns MD simulation at 300 K with NVT ensemble. During MD simulation, the position of each sp$^2$ carbon atom of graphene sheets is fixed with the harmonic constraint of 10 kcal/mol/Å$^2$ to maintain the distance between sheets. The pressure was maintained at 1 atm using the Langevin piston method with a piston period of 100 fs, a damping time constant of 50 fs, and a piston temperature of 300 K. Full electrostatics was employed using the particle-mesh Ewald method with a 1 Å grid width. Nonbonded interactions were calculated using a group-based cutoff with a switching function and updated every 10 time-step. Covalent bonds involving hydrogen were held rigid using the SHAKE algorithm, allowing a 1 fs time step. Atomic coordinates were saved every 1 ps for the trajectory analysis. The trajectories obtained from this simulation were used for calculating the translational diffusion coefficient of water in each system. All MD simulations were carried out using NAMD2 and the graphics shown in this report were prepared using VMD.

**Translational Diffusion Coefficient**

The self-diffusion coefficient $D_i$ for species i was obtained by the Einstein relation equation (2).[18,23]

$$D_i = \lim_{t \to \infty} \frac{\left\langle [r(t) - r(t_0)]^2 \right\rangle}{2dt} \quad (2)$$

, where $<>$ is the time average, $d$ the dimension of the system, and $r(t)$ the position of species i at time $t$. The self-diffusion coefficient is calculated by the slope of the least square fitted line of time versus mean squared displacement plot.

**Hydrogen Bond Lifetime**

Since the typical value of hydrogen bond lifetime is ~1 ps, we need to sample the MD trajectory with higher sampling frequency. Therefore, we performed additional 1 ns MD simulation for each system, and atomic coordinates were saved every 0.1 ps during this period. Using these samples, the lifetime of hydrogen bond ($\tau_{HB}$) of water was obtained by Luzar and Chandler introduced.[24-26] The hydrogen bond correlation function $c(t)$ is defined as

$$c(t) = \frac{\langle h(0)h(t) \rangle}{\langle h \rangle} \tag{3}$$

where $h(t) = 1$ when the given pair of donor and acceptor molecules forms hydrogen bond and 0 otherwise, and $<h>$ is average number of hydrogen bond.[25]

$$D:A \underset{k_{-1}}{\overset{k_1}{\rightleftharpoons}} D \cdots A \tag{4}$$

They adapted a kinetic model for describing the hydrogen bond formation and breaking (equation 4) where $D:A$ represents the hydrogen bond is on and $D \cdots A$ represents the hydrogen bond is off. For long-time $t$, the reactive flux is[24]

$$k(t) = -\frac{dc(t)}{dt} = -\frac{\langle \dot{h}(0)[1-h(t)] \rangle}{\langle h \rangle} \approx k_1 e^{-k_1 t} \tag{5}$$

With least square fit method, the hydrogen bond lifetime ($\tau_{HB}$) can be obtained as $\tau_{HB} = 1/k_1$. Geometric criteria are used to determine the existing hydrogen bonds: 3.5 Å cutoff of the oxygen-oxygen separation, 2.45 Å cutoff of oxygen (acceptor) and hydrogen (donor) separation, and 30° of the angle between the oxygen-oxygen vector and the covalent OH-bond.[27]

**Reorientation correlation time**

The molecular reorientational correlation times are obtained from the integration of the molecular reorientational correlation function of each water molecule.

$$\tau_l = \int_0^\infty C_l(t)dt = \int_0^\infty \langle P_l(\hat{u}(0)\cdot\hat{u}(t))\rangle dt \tag{6}$$

where $P_l(x)$ is $l$th order of Legendre polynomial and $\hat{u}(t)$ represents the unit vector bound to the water molecule at time $t$. The reorientational correlation times $\tau_1$ and $\tau_2$ have been computed for four different vectors: the OH-bond, the molecular dipole ($\mu$ or $z$-axis), HH-vector ($x$ axis) and the vector perpendicular to the molecular plane ($\perp$ or $y$-axis).

**Potential of Mean Force**

We calculated PMF of translation of water and ions through the interlayer between GO using steered molecular dynamics (SMD) simulation.[21, 28-29] During SMD simulation, the position of two GO is fixed by applying harmonic constraint while one water molecule (or an ion) is pulled with a constant velocity of 2.0 Å/ns. A harmonic constraint with a spring constant of 100 kcal/mol/Å² is used for pulling one water molecule (or an ion), and the total length of the pulling reaction coordinate is 30.0 Å. We divided the reaction coordinate into 3 consecutive sections with each section length of 10.0 Å. At each section, the system is equilibrated for 1 ns while constraining the position of two GO sheets and a water molecule (or an ion), and SMD simulation data were collected during another 5 ns. Eight independent simulations were performed at each section for constructing PMF.

The construction of PMF from SMD simulations is based on Jarzynski equality equation.

$$\Delta A = -\beta^{-1}\ln\langle\exp[-\beta W]\rangle \tag{7}$$

where $\Delta A$ is a free energy difference, $\beta$ is the product of Boltzmann factor and temperature, and $W$ is the non-equilibrium work obtained from SMD simulation. The non-equilibrium work done by the pulling force can be obtained using the following:

$$W = -kv \int_0^t dt' \left[ x(t') - x_0 - vt' \right] \tag{8}$$

where $k$ and $v$ is the force constant and velocity of pulling, $x(t')$ and $x_0$ are the reaction coordinate at $t'$ in the simulation and the initial position of the center of mass of the pulled graphene sheet. We adapted the second order cumulant expansion equation for calculating equation (7).

$$\Delta A = \langle W \rangle - \frac{\beta}{2} \left[ \langle W^2 \rangle - \langle W \rangle^2 \right] \tag{9}$$

## III. Results and Discussion

**Diffusion of water**

We monitored the structural fluctuations of each GO segment to ensure the desired interlayer distance of membrane is maintained during the MD simulations. Since all of the sp2 carbon atoms are fixed at the starting position with the harmonic constant of 10 kcal/mol/Å$^2$, we found that each GO segment maintains the planar structure, and the distance between segments is maintained well at the desired value. For example, a snapshot of GO membrane ($d = 7$ Å) obtained at $t = 10$ ns is shown in Figure 2 (A). Even though the positions of the sp2 atoms of GO are fixed, the hydroxyl and epoxy groups interact with surrounding water molecules. The diffusion coefficient ($D_w$) of water molecules in each membrane is shown in Table 2, and we found that $D_w$ increases as the interlayer distance increases. This agrees well with the previous experimental results reported by Abraham et al.[3] They found that the water flow inside the GO membrane with a larger interlayer distance is faster than that of smaller interlayer distance. We suspected that the interaction between water and GO membrane such as hydrogen bond at a shorter distance inside the membrane with shorter interlayer distance damps the translational motion of water molecules. During MD simulations, we observed that only a single water layer is formed between neighboring GO segments when $d = 7$ Å, and no water molecules penetrate the interlayer space when $d < 7$ Å. Snapshots of one water molecule between neighboring GO segments are shown in Figure 2 (B). These snapshots clearly show the hydroxyl and epoxy groups damp the translational motion of water molecules.

**Hydrogen bond lifetime**

The lifetime of two types of hydrogen bond - one between water molecules ($t_{\text{wat-wat}}$) and one between water and hydroxyl (or epoxy) group ($t_{\text{wall-wat}}$) - are listed in Table 2. As expected from the snapshots in Figure 2 (B), $t_{\text{wall-wat}}$ when $d = 7$ Å shows the longest hydrogen bond lifetime. This agrees well with the previous reports by Chandler and Luzar. Libration motion is the main reason for breaking the hydrogen bond, and the functional groups are fixed on the surface of GO membranes.

**Reorientation correlation time**

The rotational diffusion is characterized by the first and second-order correlation times of four-unit vectors bound to the water molecule (See Table 4). The rotational motion of water is anisotropic.

**Potential of Mean Force**

To investigate the free energy profile of penetration of water and ions between GO sheets, we build a system composed of two GO sheets in saltwater where the concentration of sodium and chloride ions is ~0.5 **m** that is approximately the salt concentration of seawater. Schematic views of the system are shown in Figures 4 (A) and (B). A water molecule (or ion) is pulled from $z = 16$ Å to $z = -14$ Å with a constant velocity of 2 Å/ns for calculating the free energy profile. Pulling velocity is a critical parameter for calculating the free energy difference using steered molecular dynamics simulation.[30-31] We tested different constant pulling velocities of water in the bulk phase, and we found that the free energy fluctuation is less than 0.1 kcal/mol when the pulling velocity is 2 Å/ns (See the inset of Figure 4 (C)). Therefore, we decided that the pulling velocity of 2 Å/ns is the optimum trade-off value between the accuracy and efficiency of the simulation.

When the interlayer distance $d = 13$ Å, water and ions penetrate the interlayer space without barrier and the energy fluctuation is less than 1 kcal/mol. When $d = 11$ Å, water molecules still penetrate the interlayer space with low energy fluctuations, but the free energy barriers for $Na^+$ and $Cl^-$ are increased to $\Delta A > 2$ kcal/mol. When $d = 9$ Å, water molecules also have the free energy barrier $\Delta A \approx 1.5$ kcal/mol for entering the membrane (at $z = 0$). However, once water molecules pass the entrance, the barrier inside the interlayer space is negligible. The free energy barriers of $Na^+$ and $Cl^-$ for entering the membrane are ~3 kcal/mol higher when $d = 9$. When $d = 7$ Å, the free energy barrier for ions is $\Delta A \approx 24$ kcal/mol at $z = 0$ Å, but the free energy profile of the water molecule entered the interlayer space ($z > 0$) is lowered by ~1 kcal/mol compared to that of bulk phase ($z < 0$) because of the interaction with two neighboring GO sheets. The free energy barrier for water inside the interlayer space when $d = 7$ Å shows many local minima and the minima are separated by the free energy barrier of 1~2 kcal/mol. These minima are the result of the hydrogen bond between the water molecules and the hydroxyl (or the epoxy) groups as shown in Figure 2 (B).

**Acknowledgement**

This work was supported by SEED project QEERI-S-1001.

**Table 1**. List of the system size and the number of water. All of the unit is Å. Compare with Figure 1.

| $d$ | $l_x$ | $l_y$ | $l_z$ | Number of Water |
|---|---|---|---|---|
| 7  | 28 | 52.4 | 53.8 | 1389 |
| 9  | 36 | 57.4 | 57.8 | 2679 |
| 11 | 44 | 61.2 | 61.8 | 4299 |
| 13 | 52 | 65.2 | 65.8 | 6168 |

**Table 2**. Diffusion coefficient of water (SPC/E model) inside GO membrane with the different interlayer distances between GO sheets. For comparison, we also performed an MD simulation of the water box without GO membrane and found that $D_w$ agrees well with previous experimental and computational results.

| Distance between GO, $d$ (Å) | 7 | 9 | 11 | 13 | ∞* | ∞** |
|---|---|---|---|---|---|---|
| $D_w$ (×10$^{-5}$ cm$^2$/sec) | 0.31 | 0.85 | 1.21 | 1.50 | 2.39 | 2.3 |

*Simulation

**Experiment[32]

**Table 3**. Hydrogen bond lifetime of water inside GO membrane with the different interlayer distances between GO sheets.

| $d$ (Å) | 7 | 9 | 11 | 13 | ∞* | ∞** |
|---|---|---|---|---|---|---|
| $t_{\text{wat-wat}}$ (ps) | 17.4 | 6.4 | 4.6 | 3.9 | 3.2 | ~1.0,[33-34]** 3 - 10[35-36]*** |
| $t_{\text{wall-wat}}$ (ps) | 24.2 | 6.3 | 4.6 | 3.9 | NA | NA |

*our simulation

**experimental reports in literatures

***simulation reports in literatures

**Table 4**. Rotational correlation time (ps) of water inside stacked graphene.

| $d$ (Å) | 7 | 9 | 11 | 13 | ∞* | ∞** |
|---|---|---|---|---|---|---|
| $\tau_1^{HH}$ | 12.16 | 6.92 | 5.73 | 5.34 | 4.28 | - |
| $\tau_1^{\mu}$ | 31.77 | 11.11 | 7.90 | 6.76 | 4.71 | 4.76 [37] |
| $\tau_1^{\perp}$ | 10.94 | 5.12 | 4.15 | 3.67 | 2.88 | - |
| $\tau_1^{OH}$ | 18.47 | 8.30 | 6.45 | 5.74 | 4.45 | - |
| $\tau_2^{HH}$ | 14.84 | 4.91 | 3.44 | 2.93 | 2.01 | 2.0 [38] |
| $\tau_2^{\mu}$ | 18.25 | 6.07 | 3.75 | 2.82 | 1.57 | 1.92 [37, 39] |
| $\tau_2^{\perp}$ | 32.44 | 9.20 | 4.82 | 3.24 | 1.17 | - |
| $\tau_2^{OH}$ | 13.41 | 4.79 | 3.31 | 2.69 | 1.81 | 1.95 [40-42] |

∞* Bulk water (simulation)

∞** Bulk water (experiment)

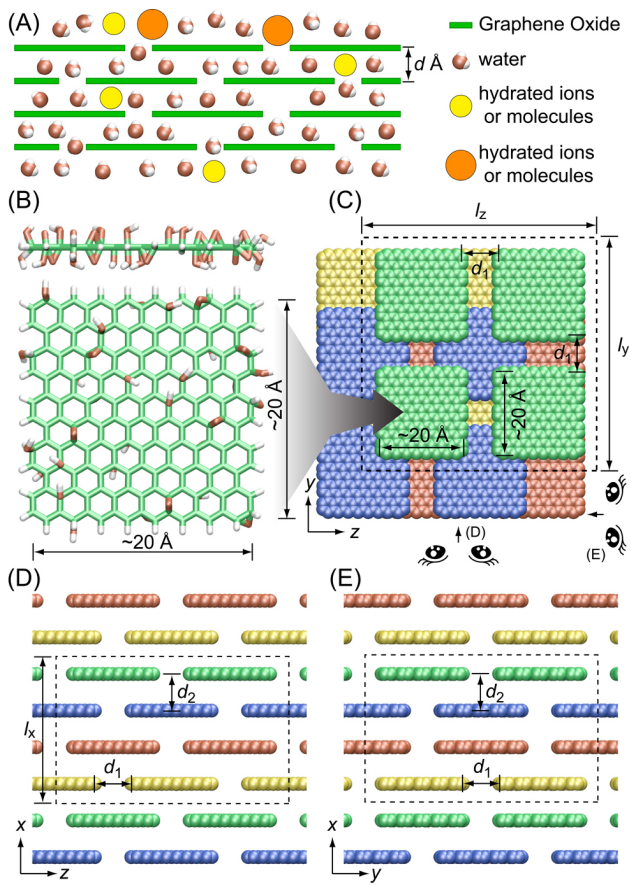

**Figure 1**. (A) Schematic representation of the penetration of water, sodium, and chloride ions through the interlayer of stacked graphene oxide membrane. (B) Snapshots of the side and top view of GO used in the simulations. The length and width of the GO are 20 Å, respectively. (C) Top view (x-y plane) of stacked GO membrane. The top layer is colored in green, the second in blue, the third in red, and the fourth in yellow. The direction of view of (D) and (E) is indicated by the schematic eye symbols. Top views of (D) x-z plane and (E) y-z plane are shown with the periodic boundaries represented by dotted lines. The distance of the horizontal ($d_1$) and vertical interlayer ($d_2$) is set to the same value, $d$ = 7, 9, 11, and 13 Å for all simulations.

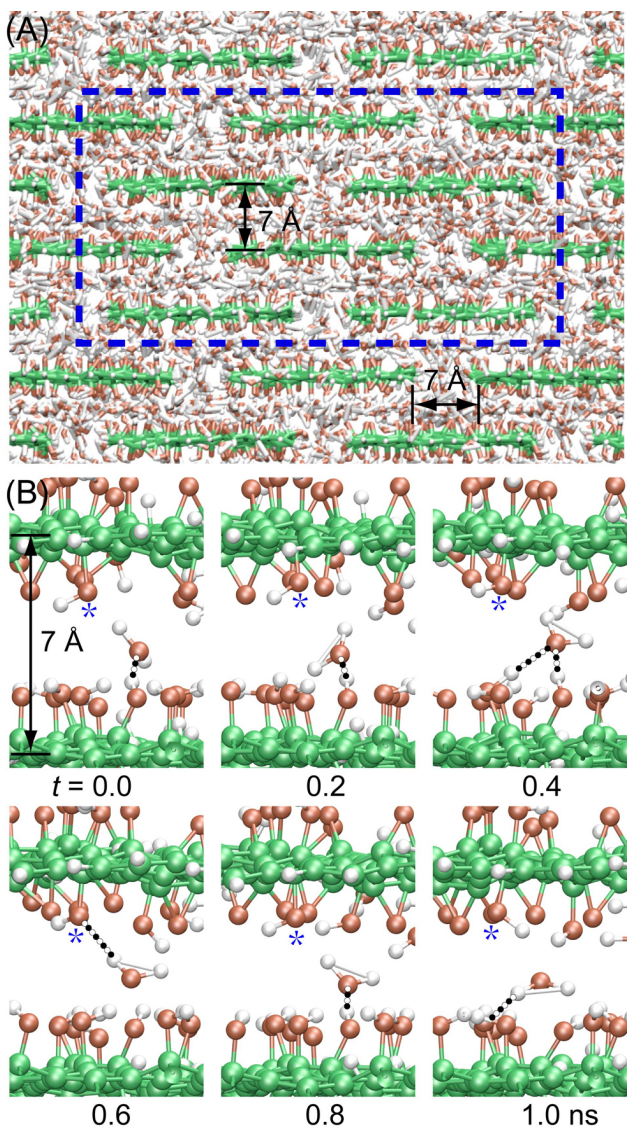

**Figure 2**. (A) A snapshot of the side view of GO membrane with $d = 7$ Å which was taken at $t = 10$ ns of MD simulation. Blue dotted lines are used for representing the periodic boundary conditions. (B) Snapshots of a selected water molecule between GO sheets when $d = 7$ from $t = 0$ to $t = 1$ ns. All other water molecules are not shown for clarity. The asterisk mark* is used for showing the relative positions of selected oxygen atoms of the same hydroxyl group of GO membrane.

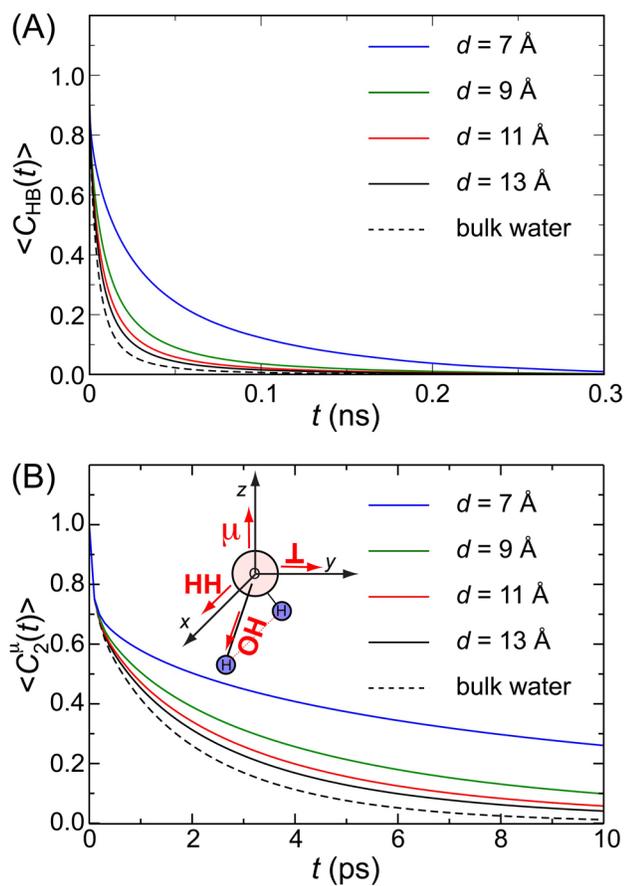

**Figure 3.** (A) The hydrogen bond autocorrelation function at different interlayer distance ($d$ = 7 ~ 13 Å). $<C_{HB}(t)>$ in bulk water is shown in dotted line for comparison. (B) Dipole rotation autocorrelation function of water when $l$ = 2. The rotation relaxation of water is slower as the interlayer distance decreases.

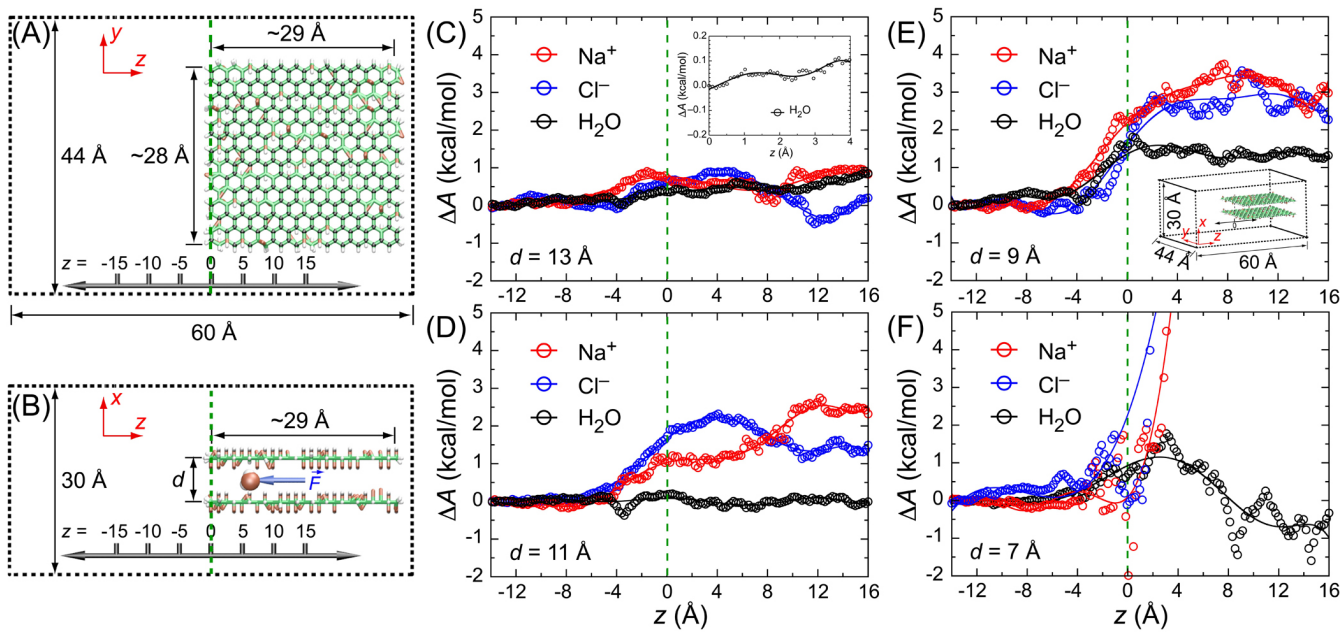

**Figure 4**. (A) Schematic top view (*y-z* plane) of the system used for calculating the free energy profile of the penetration of water and ions through the interlayer between GO sheets. The length and width of GO are ~29 Å and ~28 Å, respectively. Harmonically restrained sp2 atoms of GO are shown in black. Note that $z = 0$ is the entry position to the space between two GO sheets (indicated in green dotted lines). Water and ions are not shown for clarity. (B) A schematic side view (*x-z* plane) of the system. The water molecule (or ion) is pulled from $z = 16$ Å to $z = -14$ Å for calculating the free energy profile. Black dotted lines are used for representing the periodic boundaries. Free energy profiles of penetration of water, $Na^+$, or $Cl^-$ ions from $z = 16$ Å to $z = -14$ Å between GO sheets are shown in (C) $d = 13$ Å, (D) $d = 11$ Å, (E) $d = 9$ Å, and (F) $d = 7$ Å. Bulk water region is $z < 0$, and GO membrane region is $z > 0$.

## IV. References:

1. Nair, R. R.; Wu, H. A.; Jayaram, P. N.; Grigorieva, I. V.; Geim, A. K., Unimpeded Permeation of Water through Helium-Leak-Tight Graphene-Based Membranes. *Science* **2012**, *335*, 442-444.
2. Joshi, R. K.; Carbone, P.; Wang, F. C.; Kravets, V. G.; Su, Y.; Grigorieva, I. V.; Wu, H. A.; Geim, A. K.; Nair, R. R., Precise and Ultrafast Molecular Sieving through Graphene Oxide Membranes. *Science* **2014**, *343*, 752-754.
3. Abraham, J., et al., Tunable Sieving of Ions Using Graphene Oxide Membranes. *Nat. Nanotechnol.* **2017**, *12*, 546-+.
4. Cohen-Tanugi, D.; Grossman, J. C., Water Desalination across Nanoporous Graphene. *Nano Lett.* **2012**, *12*, 3602-3608.
5. Cicero, G.; Grossman, J. C.; Schwegler, E.; Gygi, F.; Galli, G., Water Confined in Nanotubes and between Graphene Sheets: A First Principle Study. *J. Am. Chem. Soc.* **2008**, *130*, 1871-1878.
6. Lee, O. S., Dynamic Properties of Water Confined in Graphene-Based Membrane: A Classical Molecular Dynamics Simulation Study. *Membranes* **2019**, *9*.
7. Yu, T.; Lee, O. S.; Schatz, G. C., Molecular Dynamics Simulations and Electronic Excited State Properties of a Self-Assembled Peptide Amphiphile Nanofiber with Metalloporphyrin Arrays. *J. Phys. Chem. A* **2014**, *118*, 8553-8562.
8. Bronson, J.; Lee, O. S.; Saven, J. G., Molecular Dynamics Simulation of Wsk-3, a Computationally Designed, Water-Soluble Variant of the Integral Membrane Protein Kcsa. *Biophys. J.* **2006**, *90*, 1156-1163.
9. Chun, K. M.; Kim, T. H.; Lee, O. S.; Hirose, K.; Chung, T. D.; Chung, D. S.; Kim, H., Structure-Selective Recognition by Voltammetry: Enantiomeric Determination of Amines Using Azophenolic Crowns in Aprotic Solvent. *Anal. Chem.* **2006**, *78*, 7597-7600.
10. Lee, K. H.; Lee, D. H.; Hwang, S.; Lee, O. S.; Chung, D. S.; Hong, J. I., Bowl-Shaped C-3-Symmetric Receptor with Concave Phosphine Oxide with a Remarkable Selectivity for Asparagine Derivatives. *Org. Lett.* **2003**, *5*, 1431-1433.
11. Nam, J., et al., Supramolecular Peptide Hydrogel-Based Soft Neural Interface Augments Brain Signals through a Three-Dimensional Electrical Network. *ACS Nano* **2020**, *14*, 664-675.
12. No, Y. H., et al., Nature-Inspired Construction of Two-Dimensionally Self-Assembled Peptide on Pristine Graphene. *Journal of Physical Chemistry Letters* **2017**, *8*, 3734-3739.
13. Pandey, R. P.; Rasool, K.; Rasheed, P. A.; Gomez, T.; Pasha, M.; Mansour, S. A.; Lee, O. S.; Mahmoud, K. A., One-Step Synthesis of an Antimicrobial Framework Based on Covalently Cross-Linked Chitosan/Lignosulfonate (Cs@Ls) Nanospheres. *Green Chem.* **2020**, *22*, 678-687.
14. Ulmann, P. A.; Braunschweig, A. B.; Lee, O. S.; Wiester, M. J.; Schatz, G. C.; Mirkin, C. A., Inversion of Product Selectivity in an Enzyme-Inspired Metallosupramolecular Tweezer Catalyzed Epoxidation Reaction. *Chem. Commun.* **2009**, 5121-5123.
15. Lee, O. S.; Schatz, G. C., Computational Simulations of the Interaction of Lipid Membranes with DNA-Functionalized Gold Nanoparticles. In *Biomedical Nanotechnology, Methods in Molecular Biology (Methods and Protocols)*, Hurst, S. J., Ed. Humana Press: New York, 2011; Vol. 726, pp 283-296.
16. Gao, L.; Liu, W. H.; Lee, O. S.; Dmochowski, I. J.; Saven, J. G., Xe Affinities of Water-Soluble Cryptophanes and the Role of Confined Water. *Chem. Sci.* **2015**, *6*, 7238-7248.
17. Lee, O. S.; Hwang, S.; Chung, D. S., Free Energy Perturbation and Molecular Dynamics Simulation Studies on the Enantiomeric Discrimination of Amines by Dimethyldiketopyridino-18-Crown-6. *Supramol. Chem.* **2000**, *12*, 255-272.
18. Haile, J. M., *Molecular Dynamics Simulations: Elementary Methods*; John Wiley & Sons, Inc: New York, 1992.
19. Jarzynski, C., Nonequilibrium Equality for Free Energy Differences. *Phys. Rev. Lett.* **1997**, *78*, 2690-2693.
20. Jarzynski, C., Equilibrium Free-Energy Differences from Nonequilibrium Measurements: A Master-Equation Approach. *Phys. Rev. E* **1997**, *56*, 5018-5035.
21. Park, S.; Khalili-Araghi, F.; Tajkhorshid, E.; Schulten, K., Free Energy Calculation from Steered Molecular Dynamics Simulations Using Jarzynski's Equality. *J. Chem. Phys.* **2003**, *119*, 3559-3566.
22. Al-Gaashani, R.; Zakaria, Y.; Lee, O. S.; Ponraj, J.; Kochkodan, V.; Atieh, M. A., Effects of Preparation Temperature on Production of Graphene Oxide by Novel Chemical Processing. *Ceram. Int.* **2021**, *47*, 10113-10122.
23. Einstein, A., Über Die Von Der Molekularkinetischen Theorie Der Wärme Geforderte Bewegung Von in Ruhenden Flüssigkeiten Suspendierten Teilchen. *Ann. Phys. (Berl.)* **1905**, *322*, 549.
24. Luzar, A., Resolving the Hydrogen Bond Dynamics Conundrum. *J. Chem. Phys.* **2000**, *113*, 10663-10675.
25. Luzar, A.; Chandler, D., Structure and Hydrogen-Bond Dynamics of Water-Dimethyl Sulfoxide Mixtures by Computer-Simulations. *J. Chem. Phys.* **1993**, *98*, 8160-8173.
26. Luzar, A.; Chandler, D., Hydrogen-Bond Kinetics in Liquid Water. *Nature* **1996**, *379*, 55-57.
27. Lindahl, E.; Abraham, M. J.; Hess, B.; van der Spoel, D., *GROMACS 2019.1 Manual*, February 15, 2019.
28. Park, S.; Schulten, K., Calculating Potentials of Mean Force from Steered Molecular Dynamics Simulations. *J. Chem. Phys.* **2004**, *120*, 5946-5961.

29. Lee, O. S.; Carignano, M. A., Exfoliation of Electrolyte-Intercalated Graphene: Molecular Dynamics Simulation Study. *J. Phys. Chem. C* **2015**, *119*, 19415-19422.
30. Boubeta, F. M.; Garcia, R. M. C.; Lorenzo, E. N.; Boechi, L.; Estrin, D.; Sued, M.; Arrar, M., Lessons Learned About Steered Molecular Dynamics Simulations and Free Energy Calculations. *Chem. Biol. Drug Des.* **2019**, *93*, 1129-1138.
31. Minh, D. D. L.; McCammon, J. A., Springs and Speeds in Free Energy Reconstruction from Irreversible Single-Molecule Pulling Experiments. *J. Phys. Chem. B* **2008**, *112*, 5892-5897.
32. Mark, P.; Nilsson, L., Structure and Dynamics of the TIP3P, SPC, and SPC/E Water Models at 298 K. *J. Phys. Chem. A* **2001**, *105*, 9954-9960.
33. Conde, O.; Teixeira, J., Depolarized Light-Scattering of Heavy-Water, and Hydrogen-Bond Dynamics. *Mol. Phys.* **1984**, *53*, 951-959.
34. Teixeira, J.; Bellissentfunel, M. C.; Chen, S. H., Dynamics of Water Studied by Neutron-Scattering. *J. Phys.-Condes. Matter* **1990**, *2*, SA105-SA108.
35. Antipova, M. L.; Petrenko, V. E., Hydrogen Bond Lifetime for Water in Classic and Quantum Molecular Dynamics. *Russ. J. Phys. Chem. A* **2013**, *87*, 1170-1174.
36. Voloshin, V. P.; Naberukhin, Y. I., Hydrogen Bond Lifetime Distributions in Computer-Simulated Water. *J. Struct. Chem.* **2009**, *50*, 78-89.
37. Sansom, M. S. P.; Kerr, I. D.; Breed, J.; Sankararamakrishnan, R., Water in Channel-Like Cavities: Structure and Dynamics. *Biophys. J.* **1996**, *70*, 693-702.
38. Halle, B.; Wennerstrom, H., Interpretation of Magnetic-Resonance Data from Water Nuclei in Heterogeneous Systems. *J. Chem. Phys.* **1981**, *75*, 1928-1943.
39. Krynicki, K., Proton Spin-Lattice Relaxation in Pure Water between Zero Degress C and 100 Degress C. *Physica* **1966**, *32*, 167-&.
40. Ludwig, R., NMR Relaxation Studies in Water-Alcohol Mixtures - the Water-Rich Region. *Chem. Phys.* **1995**, *195*, 329-337.
41. Struis, R.; de Bleijser, J.; Leyte, J. C., Dynamic Behavior and Some of the Molecular-Properties of Water-Molecules in Pure Water and in MgCl2 Solutions. *J. Phys. Chem.* **1987**, *91*, 1639-1645.
42. van der Maarel, J. R. C.; Lankhorst, D.; de Bleijser, J.; Leyte, J. C., On the Single-Molecule Dynamics of Water from Proton, Deuterium and O-17 Nuclear Magnetic-Relaxation. *Chem. Phys. Lett.* **1985**, *122*, 541-544.